 \documentclass[onecolumn]{emulateapj}

\usepackage{natbib}
\usepackage{graphicx}
\usepackage{rotating}
\usepackage{parskip}

\begin{document}

\title{{\it FERMI}-LAT OBSERVATIONS OF SUPERNOVA REMNANTS KESTEVEN 79}

\author{Katie Auchettl\altaffilmark{1,2,3,4}, 
        Patrick Slane\altaffilmark{1} and Daniel Castro\altaffilmark{5}}

\altaffiltext{1}{Harvard-Smithsonian Center for Astrophysics, 60 Garden Street, Cambridge, MA 02138, USA}
\altaffiltext{2}{School of Physics, Monash University, Melbourne, Victoria 3800, Australia}
\altaffiltext{3}{Monash Centre for Astrophysics, Monash University, Victoria 3800, Australia}
\altaffiltext{4}{ARC Centre of Excellence for Particle Physics at the Tera-scale, Monash University, Victoria 3800, Australia}
\altaffiltext{5}{MIT-Kavli Center for Astrophysics and Space Research, 77 Massachusetts Avenue, Cambridge, MA 02139, USA}

\begin{abstract}
In this paper we report on the detection of $\gamma$-ray emission coincident with the Galactic supernova remnant Kesteven 79 (Kes 79). We analysed approximately 52 months of data obtained with the Large Area Telescope (LAT) on board the \textit{Fermi Gamma-ray Space Telescope}. Kes 79 is thought to be interacting with adjacent molecular clouds based on the presence of strong $^{12}$CO J = 1 $\rightarrow$ 0 and HCO$^{+}$ J = 1 $\rightarrow$ 0 emission and the detection of 1720 MHz line emission towards the east of the remnant.  Acceleration of cosmic rays is expected to occur at SNR shocks, and SNRs interacting with dense molecular clouds provide a good testing ground for detecting and analysing the production of $\gamma$-rays from the decay of $\pi^0$ into two $\gamma$-ray photons. This analysis investigates $\gamma$-ray emission coincident with Kes 79, which has a detection significance of $\sim 7 \sigma$. Additionally we present an investigation of the spatial and spectral characteristics of Kes 79 using multiple archival \textit{XMM-Newton} observations of this remnant. We determine the global X-ray properties of Kes 79 and estimate the ambient density across the remnant. We also performed a similar analysis for Galactic SNR Kesteven 78 (Kes 78), but due to large uncertainities in the $\gamma$-ray background model, no conclusion can be made about an excess of GeV $\gamma$-ray associated with the remnant.
\end{abstract}


\section{INTRODUCTION}
Supernova remnants (SNRs) have long been regarded as efficient accelerators of Galactic Cosmic rays (CRs). As predicted by diffusive shock acceleration (DSA), the shock front of an SNR is expected to naturally accelerate electrons and ions with a power law momentum distribution (e.g., \citealt{2001RPPh...64..429M}), which produces non-thermal emission from the SNR. The existence of this population of particles is inferred directly from $\gamma$-rays, while the nature of the $\gamma$-ray emission and the efficiency of the particle acceleration is characterised using observations in multiple wavelengths. Non-thermal X-ray emission from a number of SNRs such as SN 1006 \citep{1995Natur.378..255K, 1998ApJ...493..375R}, Vela Jr. \citep{1998Natur.396..141A}, RX J1713.7-3946 \citep{2007Natur.449..576U} and Tycho \citep{2005ApJ...634..376W} has established that a population of relativistic electrons can be accelerated to TeV energies at SNR shocks. Ground based measurements of TeV $\gamma$-ray emission from SNRs such as  Tycho \citep{2011ApJ...730L..20A, 2012A&A...538A..81M} suggest that there is a population of particles that are being accelerated to energies approaching the knee ($E_{knee} \sim 10^{15}$ eV) of the CR energy spectrum. Even with this evidence, there has been a continuing debate in the literature  \citep[reviewed in][]{2008ARA&A..46...89R}, as to whether this $\gamma$-ray emission arises from relativistic hadrons interacting with the ambient medium (hadronic in origin), inverse Compton scattering or non-thermal bremsstrahlung emission from high-energy electrons (leptonic origin). 
\par
Gamma-rays can be produced from the decay of a neutral pion into two photons. This decay results from the interaction of relativistic ions with ambient material via proton-proton interactions \citep{1997ApJ...489..143C}. The two $\gamma$-rays, in the neutral pion's rest frame, have an energy of $\frac{1}{2} m_{\pi}c^{2}$=67.5 MeV, where $m_{\pi}$ is the rest mass of $\pi^0$ and $c$ is the speed of light. Molecular clouds (MCs) are a large source of protons, hence SNRs that are known to be interacting with dense MCs provide effective targets for detecting and analysing emission from accelerated hadrons. The interaction of an SNR's shockwave with dense molecular material is often inferred from the detection of 1720 MHz hydroxyl (OH) masers in the direction of the SNR \citep{2009ApJ...706L.270H}. Additionally, the combination of: the detection of molecular line broadening and/or asymmetric profiles, enhancement of excitation line ratios such as $J=2 \rightarrow 1/J=1 \rightarrow 0$, detection of [Fe II] or H$_{2}$ line emission due to rotational-vibrational coupling or morphological agreement of molecular features with SNR features, all provide persuasive evidence for SNR-MC interactions \citep{2013arXiv1304.5367C}.
\par 
Observations of GeV to TeV $\gamma$-rays allow one to identify $\pi^0$-decay signatures that can provide information about the parent accelerated protons. Compared to other $\gamma$-ray telescopes such as the Energetic Gamma-Ray Experiment Telescope (EGRET), the Large Area Telescope (LAT) onboard the \textit{Fermi Gamma-ray Space Telescope} has significantly improved the sensitivity and resolution in the MeV-GeV energy range, providing new opportunities for studying astrophysical objects such as SNRs. Since its launch, there have been multiple reports of the detection of GeV $\gamma$-ray emission from a large number of SNRs using data obtained from the \textit{Fermi}-LAT, such as:  IC443 \citep{2010ApJ...712..459A, 2013arXiv1302.3307T},  W44 \citep{2010Sci...327.1103A, 2013arXiv1302.3307T}, W41, MSH17-39, G337.0-0.1 \citep{2013arXiv1305.3623C}, G349.7-0.5, CTB 37A, 3C 391, G8.7-0.1 \citep{2010ApJ...717..372C} and Tycho \citep{2012ApJ...744L...2G}. Using these measurements, many authors have modeled the $\gamma$-ray spectra of these remnants to determine the mechanism behind this emission. However it has been difficult to unequivocally establish whether the $\gamma$-ray emission arises from the interaction of relativistic protons with the surrounding ambient medium or from Inverse Compton or non-thermal bremsstahlung emission. Observations of RX J1713.7-3946 using the \textit{Fermi}-LAT have suggested that the $\gamma$-ray emission is dominated by leptonic processes \citep{2010ApJ...712..287E, 2011ApJ...734...28A}. However, assuming that the SNR is interacting with a clumpy interstellar medium, \citet{2012ApJ...744...71I} concluded that the $\gamma$-rays resulted from pion decay. Gamma-ray data from the $Fermi$-LAT has also been combined with other MeV-GeV observations, such as AGILE observations of SNR W44, which strongly suggested that $\pi^0$-decay dominates the observed $\gamma$-ray emission \citep{2011ApJ...742L..30G}. This conclusion was confirmed by \citet{2013arXiv1302.3307T} who detected the characteristic pion-decay feature in the $\gamma$-ray spectra of W44 and IC443 using the $Fermi$ Large Area Telescope.
\par
Kes 79 (G33.6+0.1) is a Galactic SNR that was first discovered by \citet{1975AuJPA..37....1C} using the 408 MHz Molonglo and 5000 MHz Parkes radio continuum survey. The remnant appears to consist of two concentric incomplete radio shells with several short, bright radio filaments towards the center of the remnant  \citep{1991AJ....102..676V, 1989ApJ...336..854F}. Using single-based line interferometry over a limited range of velocities, \citet{1975A&A....45..239C} produced an H$\textsc{i}$ absorption spectrum in the direction of Kes 79, which was used to estimate a lower limit of 7 kpc for the kinematic distance. This distance estimate implies that the diameter of Kes 79 has a lower limit of 20pc. This is several times the size of a young SNR (less than a thousand years old), implying that Kes 79 is at least several thousand years old \citep{1992MNRAS.254..686G}. This conclusion was also obtained by \citet{2004ApJ...605..742S}, who derived a Sedov age of 5.9 - 7.8 kyr using a shock temperature of 0.4 - 0.7 keV.
\par
Using the Dominion Radio Astrophysical Observatory, \citet{1989MNRAS.238..737G} detected an unusually broad 1667 MHz OH absorption feature coincident with Kes 79. It was found at a position ($\alpha_{J1950}$,$\delta_{J1950}$) = ($18^{h}50^{m}10^{s}$, +$00^{\circ}35^{'}00^{"}$), with local standard of rest (LSR) velocity between +95 and +115 km s$^{-1}$.  This absorption feature coincides with a nearby molecular cloud found at the same velocity, which suggests that the remnant's shock wave is interacting with this molecular cloud. The CO survey conducted by \citet{1987ApJS...63..821S} reveals a large elongated molecular cloud at a velocity of 100 km s$^{-1}$ overlapping the eastern and southeastern region of Kes 79. The suggestion that Kes 79 is interacting with nearby molecular clouds is also supported by the detection of extended, bright $^{12}$CO J = 1 $\rightarrow$ 0 emission and strong HCO$^{+}$ J = 1 $\rightarrow$ 0 emission from the east and south-eastern regions of Kes 79 at a velocity of 105 km s$^{-1}$ by using the NRAO 12m telescope \citep{1992MNRAS.254..686G}. As the mean hydrogen density of the molecular cloud associated with Kes 79 is much less than the density required to produce significant HCO$^{+}$ emission, the authors suggested that the observed HCO$^{+}$ emission arises from the interaction with the SNR shock. Using the Parkes telescope, \citet{1997AJ....114.2058G} detected 1720 MHz line emission from Kes 79, while \citet{2008ysc..conf...41Z} reported on the detection of a 95 GHz methanol maser. However observations with the 12m Arizona Radio Observatory and VLA failed to confirm the methanol maser detection \citep{2011MmSAI..82..703F}. These observations provide additional evidence for a kinematic distance of $\sim$ 7 kpc.
\par
\citet{2004ApJ...605..742S} used a 30 ks $Chandra$ ACIS-I observation to reveal the rich structure of the X-ray emission of Kes 79, implying a complicated surrounding environment. There are many bright and faint X-ray filaments, three partial X-ray shells, a loop in the southwest and a ``protrusion" towards the northeast. Nearly all of these spatial structures have a corresponding radio structure. Using the semi-analytic, thin-shell approximation for an SNR shock crossing a density jump derived by \citet{2003ApJ...595..227C}, \citet{2004ApJ...605..742S} derived an average value of 0.36 cm$^{-3}$ for the ambient density. Using two 30 ks archival \textit{XMM} observations, \citet{2009A&A...507..841G} derived a global X-ray spectrum, which indicated dominant emission from Mg, Si, S and Fe, consistent with results derived using ASCA \citep{2000AdSpR..25..549S, 2002PASJ...54..735T}. 
\par
Kes 78 (G32.8-0.1) is a Galactic SNR that was first identified in a 408 MHz and 5000 MHz radio continuum survey by \citet{1968AuJPh..21..369K} and \citet{1975AuJPA..37....1C} respectively. It is an elongated shell type SNR which is $20' \times 10'$ in diameter and has a partially brightened non-thermal radio shell with a spectral index of $-0.5$ \citep{1992AJ....103..943K}. In the literature there are multiple estimates of the distance of Kes 78 using a number of different measurements, \citep[e.g.,][]{1975AuJPA..37....1C, 1983Ap&SS..97..287A, 1985Ap&SS.108..303G, 1998ApJ...504..761C, 2009A&A...499..789B, 2009arXiv0909.0394X}. \citet{2011ApJ...743....4Z} used H$\textsc{i}$ absorption spectra at different LSR velocities and the association of Kes 78 with a molecular cloud at an LSR velocity of $\sim$ 81 km s$^{-1}$ to derive a distance of 4.8 kpc.  \citet{1985Ap&SS.108..303G} using H$\textsc{i}$ observations, estimated the age of the remnant to be $\sim 1.2 \times 10^{5}$ yr.
\par
\citet{1998AJ....116.1323K} detected a single, 86.1 km s$^{-1}$ 1720 MHz OH maser at a position of  ($\alpha_{J2000}$, $\delta_{J2000}$) = ($18^{h}51^{m}48^{s}.04, −00^{\circ}10'35''$) using the Very Large Array (VLA). This maser is coincident with the eastern edge of Kes 78's radio shell. The detection of this maser indicates that the SNR shock-wave is interacting with the surrounding molecular clouds. Observations of $^{12}$CO reveal that Kes 78 is coincident with dense molecular clouds found towards the east of the remnant where the OH maser emission arises  \citep{2011ApJ...743....4Z}. The distribution of $^{12}$CO traces out the eastern radio shell of the remnant and it indicates that Kes 78 is expanding into a CO cavity.  Elevated $^{12}$CO $J=2 \rightarrow 1/J=1 \rightarrow 0$ ratios along the boundary of the SNR also suggest the presence of perturbations in the molecular gas due to the interaction of the SNR shock \citep{2011ApJ...743....4Z}. The kinematic distance to Kes 78 implied by the association of MCs is $4.8^{+3.1}_{-0.7}$ kpc.
\par
\citet{2011ApJ...743....4Z} detected X-rays arising from the northeastern region of the radio shell using \textit{XMM-Newton}. This emission can be modeled using a low density, under-ionised plasma with a temperature of kT $\sim$ 1.5 keV. The \textit{H.E.S.S.} collaboration reported an extended very high energy (VHE) $\gamma$-ray source, HESS J1852-000, that could be associated with the eastern edge of the remnant interacting with a nearby molecular cloud or with a previously undiscovered pulsar-wind nebula \citep{KOSACK}\footnote{\url{http://www.mpi-hd.mpg.de/hfm/HESS/pages/home/som/2011/02/}}. 
\par
In this paper, we study the $\gamma$-ray emission coincident with Kes 79 and Kes 78 and investigate the nature of this emission by modeling the broadband spectrum. Additionally we perform a spatial and spectral analysis of archival \textit{XMM} data for Kes 79 and report on the X-ray properties of this remnant. In Section 2, we describe how the Fermi-LAT data are analysed and present the results of this analysis. In Section 3, we present our method and the results of our spatial and spectral analysis of archival XMM data for Kes 79, while in Section 4 we discuss the implication of the results obtained.

\section{\textit{FERMI}-LAT OBSERVATIONS OF SNR KES 79 and KES 78}
For this study, 52 months of data, collected from August 2008 to December 2012 using the \textit{Fermi Gamma-ray Space Telescope} Large Area Telescope (\textit{Fermi}-LAT) were analysed. As previously detailed in \citet{2012arXiv1206.1896F}, only events belonging to the ``Pass7 V6" source class were selected for this study. Using this source class reduces the residual background of the data. We used the ``Pass7 version 6'' instrument response function (IRF), which were generated by using data recorded in-flight and incorporates effects that were not included in the pre-launch analysis such as accidental coincidence effects in the detectors \citep{2012arXiv1206.1896F}. For the IRFs that we used in this analysis, the systematic uncertainties in the effective area of the $Fermi$-LAT are 10\% at 100 MeV, decreasing to 5\% at 560 MeV and increasing to 10\% at 10 GeV\footnote{\url{http://fermi.gsfc.nasa.gov/ssc/data/analysis/LATcaveats.html}}. To reduce the effect of terrestrial albedo $\gamma$-rays on the data, we selected events coming from zenith angles smaller than $100^{\circ}$. To further decrease the effects of these terrestrial $\gamma$-rays, we also excluded events that were detected when the rocking angle of the LAT was greater than 52$^{\circ}$. The analysis included data from circular regions within a radius of $20^{\circ}$ centered on Kes 79 and Kes 78. The $\gamma$-ray data in the direction of Kes 79 and Kes 78 were analysed using the Fermi Science Tools v9r27p1\footnote{The Fermi Science Support Center has made publically available the Science Tools package and related documentation at \url{http://fermi.gsfc.nasa.gov/ssc}}. 
\par
To obtain morphological, positional and spectral information about the remnant, the maximum likelihood fitting technique, \textit{gtlike}, was used. Likelihood analysis was used on the \textit{Fermi}-LAT data due to the low detection rates of $\gamma$-rays and due to the large point spread function (PSF) of the \textit{Fermi}-LAT. \textit{gtlike} calculates the parameters that best fit the emission model by maximising the joint probability of the data, given a specific emission model \citep{1996ApJ...461..396M}. \textit{gtlike} uses diffuse Galactic and isotropic emission models to account for the $\gamma$-rays generated by cosmic rays interacting with background photons, the interstellar medium, the extragalactic diffuse and residual backgrounds. It also includes known $\gamma$-ray sources by placing them at fixed positions and calculating the background contribution from these sources to the data. The $\gamma$-ray emission from the Milky Way is described by the mapcube file \texttt{gal\_2yearp7v6\_v0.fits}, while the table \texttt{iso\_p7v6source.txt} models the isotropic emission arising from extragalactic diffuse and residual backgrounds\footnote{The most up to date \texttt{gal\_2yearp7v6\_v0.fits} and \texttt{iso\_p7v6source.txt} where obtained from \url{http://fermi.gsfc.nasa.gov/ssc/data/access/lat/BackgroundModels.html}}. 
\subsection{Spatial Analysis}
Gamma-ray data with energy ranging from 2 to 200 GeV converted in the \textit{front} section of the LAT were selected to determine the spatial characteristics of the $\gamma$-ray emission from Kes 79 and Kes 78. These events were chosen as they improve the angular resolution of the data. The 1$\sigma$ containment radius angle for \textit{front}-selected photons in this energy band is $\leq 0.3^{\circ}$. Test statistic (TS) maps constructed using \textit{gttsmap} allow one to determine the detection significance as well as the position and extent of the source.  The test statistic is defined as the logarithmic ratio of the likelihood of a point source being found at a given position on a spatial grid, to the likelihood of the model without the additional source, $2\log(L_{ps}/L_{null})$. The image resolution of these TS maps is defined by the size of the grid used for this analysis; we set the grid to be $0.05^{\circ}$.
\par
\begin{figure}[t!]
\includegraphics[width=0.5\textwidth]{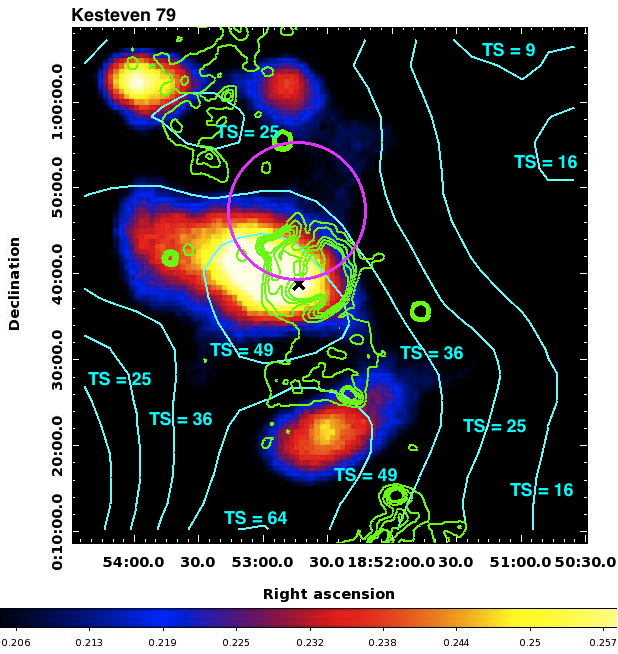}
\includegraphics[width=0.5\textwidth]{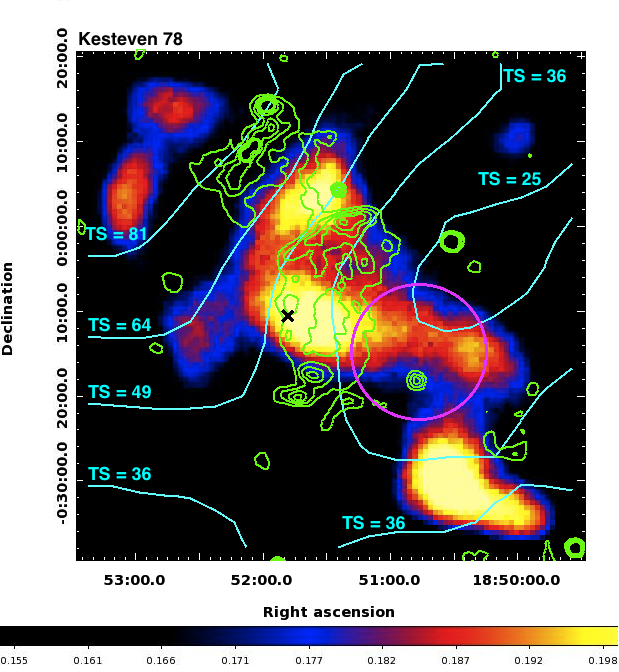}
\caption{\textit{Fermi}-LAT count maps of events with energy of 2 to 200 GeV in a region $1^{\circ} \times 1^{\circ}$ surrounding Kes 79 and Kes 78 respectively  (units are \textit{counts $degree^{-2}$}). The pixel binning is $0.01^{\circ}$ and the maps are smoothed with Gaussians of width 0.2 arcminutes. The magenta circles corresponds to the \textit{Fermi}-LAT sources, 2FGL J1852.7+0047c and 2FGL J1850.7-0014c, located near Kes 79 and Kes 78 respectively. The radio emission observed during the VLA Galactic plane survey are overlaid as the green contours \citep{2006AJ....132.1158S}, while the $Fermi$-LAT TS map contours are shown in cyan. The square root of the TS value is approximately equal to the detection significance. The position of the 1667 MHz OH absorption feature for Kes 79 and the position of the 1720 MHz OH maser for Kes 78 are indicated by the black crosses. The scale corresponds to the number of \textit{counts degree$^{-2}$}. \label{fig:count}} 
\end{figure}

To analyse the surrounding neighbourhood of the remnant, we generated count maps of $1^{\circ} \times 1^{\circ}$ regions centered around Kes 79 and Kes 78. The count maps were smoothed by a Gaussian of width similar to the PSF of the events selected and are presented in Figure \ref{fig:count}. The radio emission from Kes 79 and Kes 78 as observed during the VLA Galactic plane survey are overlaid as the green contours \citep{2006AJ....132.1158S}. Possibly associated with Kes 79 is an unresolved $\gamma$-ray source designated as 2FGL J1852.7+0047c in the \textit{Fermi}-LAT Second Source Catalogue, while possibly associated with Kes 78 is an unresolved $\gamma$-ray source  designated as 2FGL J1850.7-0014c. In the Second Source Catalogue, these two \textit{Fermi}-LAT sources are characterised by a photon power law index of $\Gamma = 2.53 \pm 0.21$ and $\Gamma = 2.85 \pm 0.18$ respectively.  The photon flux of these sources over an energy of 1 GeV - 100 GeV is $(1.8 \pm 0.5)  \times 10^{-9}$ photons cm$^{-2}$ s$^{-1}$ and $(4.3 \pm 0.9)  \times 10^{-9}$ photons cm$^{-2}$ s$^{-1}$ respectively.  The position and extent of each of these $\gamma$-ray sources are plotted as a magenta circle, in Figure \ref{fig:count}.
\par
To determine the significance of the detection of the $\gamma$-ray emission from Kes 79 and Kes 78, we generated a TS map for each remnant. For Kes 79, we calculated the TS map over an energy range of 0.2 - 2 GeV and obtained a detection significance of $\sim 7 \sigma$. The TS map is presented as the cyan contours overlaid in Figure \ref{fig:count}a. The TS contours suggest that the $\gamma$-ray emission in this region is associated with Kes 79 . For Kes 78 we calculated a TS map over an energy range of 1 GeV - 3.2 GeV and overlaid the TS map as cyan contours in Figure \ref{fig:count}b. 
We select this range, as the resolution of the $Fermi$-LAT below 1 GeV is significantly affected by the large uncertainty in the Galactic diffuse emission model. The TS contours do not show a concentration (or peak) of flux at the centroid of Kes 78, this is in contrast to the significant $\gamma$-ray emission seen in the region centered on Kes 79 (Figure \ref{fig:count}a). The detection significance of $\sim 5 - 6 \sigma$ in the region of Kes 78 is equal to or less than the detection significance seen everywhere in the field of view indicating that the observed $\gamma$-ray emission is dominated by uncertainities in the background model or by $\gamma$-ray emission spilling over from nearby sources. Thus it is not possible to claim an excess of GeV $\gamma$-ray emission associated with the remnant.

\subsection{Spectral Analysis}
\begin{figure}[t!]
\begin{center}
\includegraphics[width=0.75\textwidth]{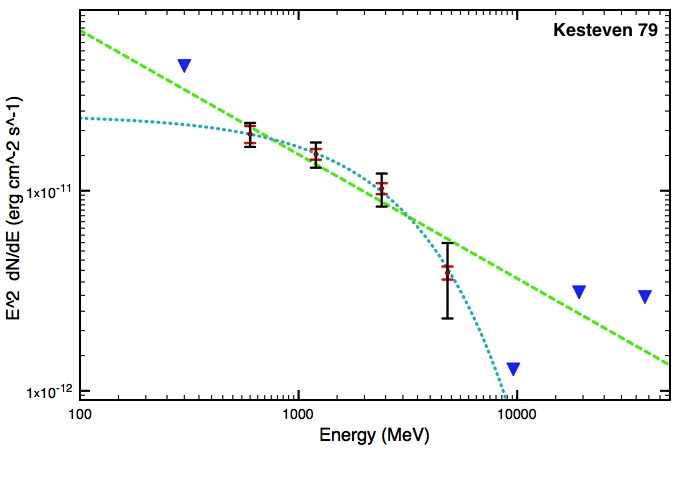}
\caption{The \textit{Fermi}-LAT gamma-ray spectrum of Kesteven 79. The statistical errors are plotted in black,  systematic errors are plotted in red and upper limits are plotted in blue. The simple power law models and exponential cut-off model described by the parameters in Table \ref{table:1} are shown as the green dotted line and the cyan dashed line respectively.  \label{fig:gamma}}
\end{center}
\end{figure}


The spectral energy distribution (SED) characteristics of the \textit{Fermi}-LAT emission coincident with Kes 79 is produced by using events with an energy of 0.2 - 200 GeV. This energy range is selected for multiple reasons. These include: avoiding source confusion; avoiding the large flux uncertainties that arise below 0.2 GeV due to limitations of the Galactic diffuse model; and reducing the influence that the rapidly changing effective area of the instrument at low energies has on the data. We perform the spectral analysis by modeling the flux at each energy bin and then estimate the parameters that best fit the data using the maximum likelihood technique \textit{gtlike}. We also included in the likelihood fits background sources from the 24-month \textit{Fermi}-LAT Second Source Catalogue \citep{2012ApJS..199...31N} that are found within the 20$^\circ$ circular region centered on Kes 79\footnote{The data for these sources is made available by the Fermi Science Support Center and is found at \url{$http://fermi.gsfc.nasa.gov/ssc/data/access/lat/2yr\_catalog/$}}. 
For Kes 79, all evident background sources were identified in the \textit{Fermi}-LAT Second Source Catalogue and the associated parameters from the catalogue were used. 
\par
For each energy bin, the normalisation of the diffuse Galactic component was left free to account for correlations between close-by sources. In addition to using the statistical uncertainties from the likelihood analysis, the systematic uncertainty associated with the Galactic diffuse emission was also considered by artificially altering the nxormalisation of the Galactic background by $\pm 6$\% from the best-fit value at each energy bin. This method is modeled from similar treatments presented in \citet{2010ApJ...717..372C} and \citet{2012ApJ...756...88C}.
\par
The $\gamma$-ray spectrum of the \textit{Fermi}-LAT sources coincident with Kes 79 is shown in Figure \ref{fig:gamma}. Upper limits are plotted as blue triangles. The $\gamma$-ray spectrum of Kes 79 can be fitted using a simple power law with a spectral index of 2.6$\pm$0.1, while an exponential cut-off power law with $E_{cut}$=2.7$\pm$ 0.6 GeV and a spectral index of 2.00$\pm$0.4, significantly improves the fit.  Assuming a distance of 7 kpc to Kes 79, the luminosity of this $\gamma$-ray source for an energy range of 0.1 - 100 GeV is 6.1$\times$10$^{35}$ erg s$^{-1}$. 
Both the simple power law and exponential cut-off models for Kes 79 are shown in Figure \ref{fig:gamma} as the green dotted and cyan dashed plots respectively. The parameters which define these best fits, along with the detection significance and positional information of Kes 79 are summarised in Table \ref{table:1}. 
\begin{table}[t!]
\begin{center}
      \caption{The spatial and spectral fit parameters for the Kes 79 {\it Fermi}-LAT data. \label{table:1}}
\begin{tabular}{lccccccccccc}
\hline
  & \multicolumn{4}{c}{\textit{Spatial}} & \multicolumn{6}{c}{\textit{Spectral Fit}} \\
 \cline{2-5} 
   \cline{7-11} 
 ~      & R.A.   & Decl.  &  ~   & \textit{F(0.1-100GeV)}                    && ~                      &~ &  \\ 
        Name   & (deg)  & (deg)  &  ~   & \tiny{(10$^{-7}$ photons GeV cm$^{-2}$ s$^{-1}$)} && $\Gamma_{pwl}$& $\chi_{pwl}^{2}$(dof)& $\Gamma_{cut}$& $E_{cut}$ (GeV)& $\chi_{exp}^{2}$(dof)& TS value  \\ 
\hline
     Kes 79 & 283.121 & +0.645 &   & $0.63^{+0.70}_{-0.30}$         &           & 2.62$\pm$0.12  & 2.87(2)&$2.00\pm 0.38$ &$2.71\pm 0.64$& 0.27(1)& $\sim 7\sigma$    \\
\hline
\end{tabular}
\end{center}
\vspace{0.5cm}
\end{table}


\section{XMM-NEWTON OBSERVATIONS AND ANALYSIS}

For Kes 79 we analysed 21 archival \textit{XMM-Newton} observations which were found within a radius of 12 arcmins of the centroid of Kes 79 (see Table \ref{table:2}). These observations were performed over a period of 5 years and have an effective exposure time of $\sim$447 ks and $\sim$456 ks for the MOS 1 and MOS 2 detector respectively. All \textit{XMM-Newton} observations simultaneously acquired EPIC MOS and EPIC-pn observations, with the pn detector operating in ``small window" mode, while the MOS cameras operated in ``full frame" mode. For all observations, the full SNR was included in the field of view of the MOS cameras, while only part of the remnant was detected by EPIC-pn due to it operating in ``small window" mode. Thus for this analysis, we only used data from the MOS cameras. 
\par
\citet{2009A&A...507..841G} used the \textit{XMM-Newton} observations 204970201 and 204970301 to compare the morphology of the X-ray emission and 324 MHz VLA observations of Kes 79, while \citet{2010ApJ...709..436H} used the EPIC-pn observations of 16 of the \textit{XMM-Newton} observations listed in Table \ref{table:2} to perform a dedicated series of timing observations of PSR J1852+0040 found in Kes 79. We re-analysed all 21 archival EPIC MOS observations previously presented in the literature using the XMM-Newton Science Analysis System (SAS) version 12.0.1 and the most up-to-date calibration files\footnote{Documentation related to the SAS package, as well as its download is distributed by XMM-Newton Science Operations Center at \url{http://xmm.esac.esa.int/sas/} }. All analyses were completed by starting from the observational data files (ODFs). For the MOS cameras, we selected single to quadruple patterned events. For both spectral and imaging analysis, all event files were filtered such that all flagged events were removed and periods of high background and/or photon flare contamination were removed as suggested in the current SAS analysis threads and XMM-Newton Users Handbook\footnote{\url{http://xmm.esac.esa.int/external/xmm\_user\_support/documentation/uhb/index.html}}. We generated count rate histograms for all observations with energies above 10 keV to determine the time intervals where the emission from the MOS cameras was affected by periods of high background and/or photon flares. The effective exposures of the MOS 1 and MOS 2 observations after filtering are shown Table \ref{table:2}. 
\begin{center}
\begin{table}[t!]

      \caption{Archival \textit{XMM-Newton} observations of Kes 79 and their corresponding total and effective exposure times as used in our analysis.\label{table:2}}
 
\begin{tabular}{lccccccccccc}
\hline
& R.A. & Decl. & & & MOS 1  & MOS 2 \\
Obs. ID & (J2000) & (J2000) & Observation Date & Exposure (ks) & Good Exposure (ks) & Good Exposure (ks)  \\
\hline
204970201 & 18 52 35.54 & +00 39 40.4 & 10/18/04 & 31.4 & 28.7 & 30.3 \\
204970301 & 18 52 35.65 & +00 39 36.4 & 10/23/04 & 31.4 & 30.9 & 30.5 \\
400390201 & 18 52 34.90 & +00 39 42.4 & 10/08/06 & 30.4 & 28.7 & 29.4 \\
400390301 & 18 52 42.56 & +00 40 52.2 & 03/20/07 & 34.5 & 34 & 33.6 \\
550670201 & 18 52 34.72 & +00 39 48.7 & 09/19/08 & 26.9 & 21.6 & 21.8 \\
550670301 & 18 52 34.61 & +00 39 46.3 & 09/21/08  & 35.4 & 30.4 & 30.5 \\
550670401 & 18 52 34.64 & +00 39 48.1 & 09/23/08 & 40.4 & 36.1 & 35.5 \\
550670501 & 18 52 34.68 & +00 39 43.0 & 09/29/08  & 34.1 & 32.7 & 33.1 \\
550670601 & 18 52 34.82 & +00 39 43.7 & 10/10/08  & 40.4 & 34.7 & 34.1 \\
550670901 & 18 52 42.67 & +00 40 48.5 & 03/17/09 & 26.9 & 25 & 25.4 \\
550671001 & 18 52 42.53 & +00 40 49.0 & 03/16/09  & 27.9 & 23.5 & 24.5 \\
550671101 & 18 52 42.31 & +00 40 50.4 & 03/25/09  & 23.0 & 19.0 & 18.7 \\
550671201 & 18 52 42.49 & +00 40 52.2 & 03/23/09  & 27.9 & 17.7 & 19.4 \\
550671301 & 18 52 42.24 & +00 40 58.1 & 04/04/09  & 26.9 & 22.2 & 22.7 \\
550671401 & 18 52 42.67 & +00 40 48.4 & 03/17/09  & 5.93 & 5.4 & 5.3 \\
550671501 & 18 52 42.28 & +00 40 52.0 & 03/25/09 & 6.42 & 0.0 & 4.3 \\
550671601 & 18 52 42.49 & +00 40 52.0 & 03/23/09 & 5.63 & 5.0 & 5.1 \\
550671701 & 18 52 42.24 & +00 40 57.9 & 04/04/09 & 4.43 & 3.7 & 3.8 \\
550671801 & 18 52 41.92 & +00 41 02.9 & 04/22/09  & 28.9 & 25.4 & 25 \\
550671901 & 18 52 42.10 & +00 40 58.4 & 04/10/09  & 31.4 & 18.1 & 19.5 \\
550672001 & 18 52 42.10 & +00 40 58.3 & 04/10/09  & 5.13 & 3.8 & 3.6 \\
\hline
\end{tabular}
\vspace{0.5cm}
\end{table}
\end{center}
\par
\subsection{Imaging}
To analyse the morphology of Kes 79, we produced an image of the entire remnant by combining all 21 filtered observations of MOS 1 and MOS 2 using the SAS task \textit{emosaic}. This task produces a single exposure-corrected intensity image of the SNR. For each camera, we generated maps in the energy bands 0.5 - 1.195 keV, 1.195 - 1.99 keV and 1.99 - 7.00 keV.
\par
\begin{figure}[t!]
\includegraphics[width=0.5\textwidth]{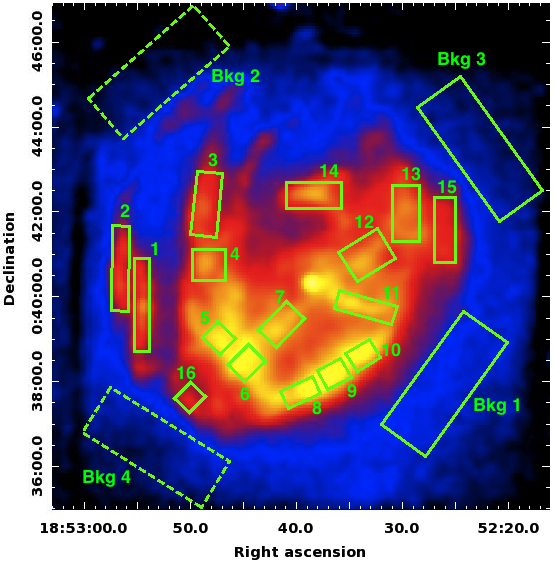}
\includegraphics[width=0.5\textwidth]{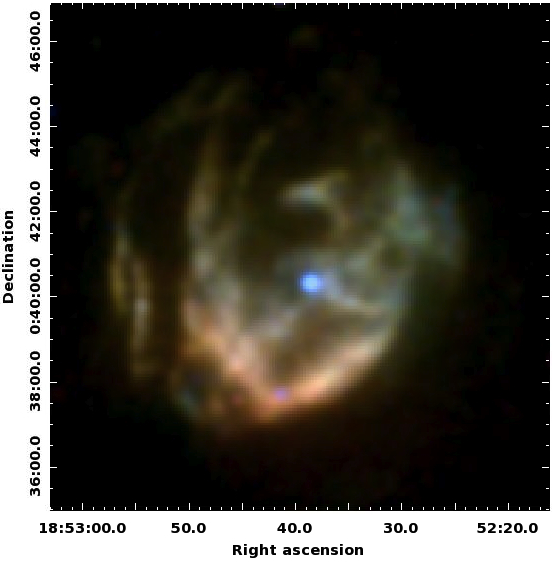}
\caption{a) A 0.5 - 7 keV exposure corrected image of the X-ray emission from Kes 79. This was created using the \textit{XMM-Newton} SAS task \textit{emosaic}. The image has been smoothed with a Gaussian function of width 3". Overlaid in green are the regions we use for spectral extraction, described in Section 3.2, and the four background regions we used. b) Exposure corrected RGB image of SNR Kes 79 created using EPIC MOS1 and MOS2 data. Red corresponds to 0.5 - 1.195 keV energy band, green to 1.195 - 1.99 keV and blue to 1.99 - 7.00 keV. The RGB image has been smoothed with a Gaussian function of width 3".  The relative intensity levels of the different colour images have been adjusted to highlight spectral structures of the emission.\label{fig:xraystuff}}
\end{figure}

The \textit{XMM-Newton} EPIC MOS mosaic image of Kes 79 in the 0.5 - 7.0 keV band is presented in Figure \ref{fig:xraystuff}a. The X-ray emission observed with \textit{XMM-Newton} corresponds well to the rich spatial structure observed by \citet{2004ApJ...605..742S} using 29.3 ks of \textit{Chandra} data. A large majority of the emission comes from the center of the remnant, with faint X-ray emission found towards the northeast. There are multiple faint and bright X-ray filaments and evidence of multiple partial shells highlighted by bright X-ray emission towards the south. Also observable in this image is the X-ray emission from the pulsar PSR J1852+0040. Our mosaic image mimics the 0.5 - 5 keV \textit{XMM} image of Kes 79 by \citet{2009A&A...507..841G}.
\par
With the aim of analysing any possible spectral variation in Kes 79, we also created an RGB image of the remnant which is found in Figure \ref{fig:xraystuff}b. We generated this image by combining images of events from the MOS 1 and MOS 2 cameras in the energy bands 0.5 - 1.195 keV (red), 1.195 - 1.99 keV (green) and 1.99 - 7.00 keV (blue). The relative intensities of the colour images have been adjusted in the RGB representation to highlight all structures. This image reveals a very filamentary and clumpy morphology, with the southern shell producing a significant amount of soft X-rays. The pulsar, which has a hard spectrum, is seen as the bright blue dot in the center of the image. 
\subsection{Spectroscopy}

The aim of our spectral analysis of Kes 79 is to perform a spatially resolved, spectral study of the remnant. To do this, we analyse the variation in the spectral parameters in different regions of the remnant.
\par
For our analysis, we used \textit{XMM} observations that had a good exposure time of 10 ks or more, as the event files with less than this amount produced spectra with less than 100 counts. The X-ray spectra were extracted from the filtered and cleaned event files using the SAS task \textit{evselect}. The task \textit{arfgen} and \textit{rmfgen} were used to generate a spectral response and effective area file for each extracted region.  
\par
The background of \textit{XMM-Newton} consists of the Cosmic X-ray background and a time variable non X-ray background resulting from electronic noise, solar protons and cosmic rays interacting with the detector \citep{2008A&A...478..575K}. Above $\sim$ 5 keV, the X-ray spectrum for these observations is dominated by background. We accounted for the background of each observation by selecting four source-free regions as shown in Figure \ref{fig:xraystuff}a. Four background regions were selected to ensure that any gaps due to mosaicking would not affect the spectra obtained.
\par
The combined spectra were analysed with the X-ray software XSPEC version 12.8.0, over an energy range of 0.5 - 10 keV. Each energy bin was required to have a minimum of 25 counts and the reduced $\chi^{2}$ statistic was used as the best-fit model discriminator. To investigate the emission of the remnant we used the non-equilibrium ionisation (NEI) plasma emission models, PSHOCK and VPSHOCK \citep{2001ApJ...548..820B}. These models describe a plane-parallel shock front interacting with a plasma and are characterised by a constant electron temperature (kT) and an upper and lower limit on the ionisation timescale ($\tau_{l} =n_{e}t_{l}$ and $\tau_{u} =n_{e}t_{u}$). The absorption, upper limit of the ionisation timescale, temperature, normalisation of both models, and the abundances of Ne, Mg, Si, and S, were left as free parameters. All other elemental abundances were frozen to the solar values of \citet{2000ApJ...542..914W} once we verified their variation did not significantly improve the model fit of the data. As these NEI models do not include atomic data for the Argon emission, an additional Gaussian component with a width of 0.01 keV and line energy of 3.1396 keV was added to account for this feature in the spectra fitting. The normalisation of this additional component was left free. 
\newpage
\subsubsection{Spectral properties of individual regions}
\begin{sidewaystable}[ph!]
\begin{center}
\caption{Results from fitting the two component model VPSHOCK+PSHOCK model and the XSPEC absorption model TBABS the 16 different regions defined in Figure \ref{fig:xraystuff}a, with the 90\% confidence ranges. \label{table:individual}}
\begin{tiny}
\begin{tabular}{cccccccccccccccc}
\hline
\multicolumn{3}{c}{\textit{TBABS}} & \multicolumn{2}{c}{\textit{Soft Component (PSHOCK)}} & \multicolumn{7}{c}{\textit{Hard Component (VPSHOCK)}}\\
\cline{2-12}
    Region&nH &kT & $\tau_{u}$&Norm $cm^{-5}$  &kT &$\tau$ &Norm $cm^{-5}$ &Ne&Mg&Si&S                      & $\chi^{2}/(dof)$ \\ 
    &($10^{22}$ cm$^{2}$)&(keV)& ($10^{12} s/cm^{3}$)& ($10^{-2}$) & (keV)& ($10^{11} s/cm^{3}$)& ($10^{-3}$)&&&&                      &  \\ \hline
    1      & $2.43 \pm 0.05$        & $0.19 \pm 0.01$          & $1$    & $5.94^{+2.5}_{-2.0}$ & $0.55^{+0.03}_{-0.01}$ & $2.76^{+0.6}_{-0.7}$   & $3.61^{+0.1}_{-0.5}$   & $-$                  & $1.29 \pm 0.8$         & $2.31^{+0.1}_{-0.2}$   & $2.50 \pm 0.3$         & $1.19$           \\
    2      & $2.51^{+0.06}_{-0.03}$ & $0.20 \pm 0.02$          & $1$    & $3.14^{+2.5}_{-1.3}$ & $0.49 \pm 0.03$        & $2.61^{+0.9}_{-0.6}$   & $3.41^{+1.0}_{-1.1}$   & $-$                  & $1.42 \pm 0.1$         & $2.69 \pm 0.3$         & $2.83^{+0.5}_{-0.4}$ & $1.10$           \\
    3      & $2.48^{+0.04}_{-0.02}$ & $0.25 \pm 0.01$          & $1$    & $3.77^{+1.2}_{-0.3}$ & $0.78^{+0.02}_{-0.04}$ & $1.56 \pm 0.2$         & $2.71^{+0.6}_{-0.2}$   & $-$                  & $1.52^{+0.08}_{-0.1}$  & $2.53^{+0.2}_{-0.1}$   & $2.53 \pm 0.2$         & $1.24$           \\
    4      & $2.14^{+0.02}_{-0.03}$ & $0.22 \pm 0.01$          & $1$    & $3.62^{+0.3}_{-0.4}$ & $0.69^{+0.01}_{-0.02}$ & $1.99^{+0.3}_{-0.1}$   & $3.51^{+0.4}_{-0.2}$   & $-$                  & $1.44^{+0.06}_{-0.05}$ & $2.18^{+0.09}_{-0.06}$ & $2.33 \pm 0.12$        & $1.32$           \\
    5      & $2.12^{+0.02}_{-0.01}$ & $0.23 \pm 0.005$         & $1$    & $4.00^{+1.2}_{-0.6}$ & $0.75 \pm 0.02$        & $1.74 \pm 0.2$         & $3.89 \pm 0.3$         & $-$                  & $1.39^{+0.05}_{-0.04}$ & $2.11^{+0.06}_{-0.05}$ & $2.06 \pm 0.1$         & $1.34$           \\
    6      & $2.27 \pm 0.03$        & $0.25^{+0.009}_{-0.003}$ & $1$    & $3.72^{+0.7}_{-0.4}$ & $0.80^{+0.03}_{-0.01}$ & $2.44^{+0.2}_{-0.3}$   & $4.40^{+0.09}_{-0.64}$ & $-$                  & $1.40 \pm 0.04$        & $1.94 \pm 0.06$        & $1.88^{+0.12}_{-0.08}$ & $1.35$           \\
    7      & $2.08^{+0.03}_{-0.04}$ & $0.26 \pm 0.01$          & $1$    & $3.67^{+0.4}_{-0.6}$ & $1.05^{+0.05}_{-0.09}$ & $9.25^{+0.16}_{-0.07}$ & $1.34^{+0.6}_{-0.2}$   & $1.79 \pm 0.27$      & $1.84^{+0.13}_{-0.19}$ & $3.23^{+0.44}_{-0.48}$ & $2.87^{+0.30}_{-0.46}$ & $1.43$           \\
    8      & $2.09^{+0.02}_{-0.04}$ & $0.25^{+0.02}_{-0.01}$   & $1$    & $2.35^{+0.4}_{-0.6}$ & $0.67^{+0.03}_{-0.02}$ & $3.31^{+0.6}_{-0.5}$   & $4.08^{+0.6}_{-0.4}$   & $-$                  & $1.41^{+0.07}_{-0.05}$ & $2.07 \pm 0.1$         & $2.25 \pm 0.2$         & $1.38$           \\
    9      & $2.30 \pm 0.03$        & $0.23 \pm 0.01$        & $1$    & $4.09 \pm 0.1$       & $0.73^{+0.04}_{-0.02}$ & $2.34^{+0.2}_{-0.3}$   & $3.15^{+0.3}_{-0.6}$   & $1.28^{+0.3}_{-0.1}$ & $1.56^{+0.2}_{-0.1}$   & $2.35^{+0.3}_{-0.1}$   & $2.33^{+0.3}_{-0.1}$   & $1.36$           \\
    10   & $2.61^{+0.02}_{-0.04}$        & $0.26^{+0.003}_{-0.005}$          & $1$    & $5.74^{+0.3}_{-0.2}$       & $0.90 \pm 0.2$         & $1.57 \pm 0.10$         & $4.54^{+0.3}_{-0.2}$         & $0.94 \pm 0.1$       & $1.28^{+0.06}_{-0.07}$         & $2.03^{+0.09}_{-0.08}$         & $1.97 \pm 0.1$         & $1.34$           \\
    11     & $2.62^{+0.03}_{-0.02}$ & $0.25^{+0.003}_{-0.004}$ & $1$    & $7.78^{+0.7}_{-2.2}$ & $0.80^{+0.01}_{-0.03}$ & $1.59 \pm 0.1$         & $5.14^{+0.7}_{-0.6}$   & $1.18^{+0.1}_{-0.2}$ & $1.70^{+0.05}_{-0.06}$ & $2.72 \pm 0.2$         & $2.42 \pm 0.1$         & $1.34$           \\
    12     & $2.75^{+0.03}_{-0.04}$ & $0.23 \pm 0.01$          & $1$    & $8.65^{+3.0}_{-1.8}$ & $0.81^{+0.03}_{-0.05}$ & $0.96^{+0.06}_{-0.05}$ & $3.12^{+0.6}_{-0.5}$   & $-$                  & $2.09^{+0.1}_{-0.2}$   & $4.39^{+0.2}_{-0.4}$   & $3.63^{+0.3}_{-0.2}$   & $1.53$           \\
    13     & $2.50^{+0.06}_{-0.04}$ & $0.23 \pm 0.02$          & $1$    & $3.03^{+1.6}_{-0.7}$ & $0.61 \pm 0.03$        & $3.66^{+0.7}_{-1.1}$   & $4.71 \pm 0.8$         & $0.84^{+0.2}_{-0.1}$ & $1.34 \pm 0.06$        & $2.01^{+0.2}_{-0.1}$   & $2.17^{+0.2}_{-0.1}$   & $1.18$           \\
    14     & $2.53^{+0.04}_{-0.07}$ & $0.23 \pm 0.02$          & $1$    & $3.49 \pm 1.2$       & $0.65^{+0.03}_{-0.04}$ & $2.92 \pm 0.8$         & $4.14 \pm 0.9$         & $0.91 \pm 0.2$       & $1.35 \pm 0.1$         & $2.04 \pm 0.2$         & $2.13 \pm 0.2$         & $1.22$           \\
    15     & $2.45 \pm 0.1$         & $0.22^{+0.03}_{-0.02}$   & $1$    & $1.56 \pm 0.1$       & $1.36 \pm 0.3$         & $0.41 \pm 0.09$        & $0.17 \pm 0.01$        & $-$                  & $2.38^{+0.5}_{-0.6}$   & $5.81^{+2.3}_{-1.7}$   & $4.14^{+1.6}_{-1.1}$   & $1.09$           \\
    16     & $2.51^{+0.05}_{-0.06}$ & $0.24 \pm 0.02$          & $1$    & $3.08^{+1.6}_{-1.0}$ & $0.62 \pm 0.03$        & $3.60^{+1.4}_{-0.9}$   & $4.65^{+0.9}_{-0.5}$   & $0.56 \pm 0.2$       & $1.34^{+0.09}_{-0.08}$ & $2.04^{+0.3}_{-0.2}$   & $2.21 \pm 0.2$         & $1.18$           \\
    \hline

\end{tabular}
\end{tiny}
\end{center}
\end{sidewaystable}

To investigate the spectral variation across Kes 79, we perform a spatially resolved spectral analysis by extracting X-ray spectra from 16 regions shown in green in Figure \ref{fig:xraystuff}a. These regions were selected as they correspond to variations in the colour distribution in the RGB image or variation in the flux as observed in the X-ray image. To increase the signal-to-noise ratio, we combined the X-ray spectra extracted for all 16 MOS1 and MOS2 observations with good time events of 10 ks or greater, producing combined MOS1 and combined MOS2 X-ray spectra, for each region. This was done by using the HEASOFT tasks: \textit{mathpha, fparkey, marfrmf} and \textit{addrmf}. The combined spectra for all regions were grouped using the FTOOLS command \textit{grppha}.
\par
Initially the spectra from all extracted regions were fitted using a single component model (VPSHOCK) and the XSPEC absorption model TBABS \citep{2000ApJ...542..914W}. By itself, TBABS*VPSHOCK was able to account reasonably well for the X-ray emission below $\sim$~ 3 keV, but it failed to fit the higher energy component of the spectrum and as a consequence the fits resulted in relatively large reduced chi-squared values ($\chi^{2}/(dof) \textgreater 1.80$). Subsequently we fitted the spectra of each region with a two component plasma model, VPSHOCK+PSHOCK, and TBABS. The parameters of the best-fit models for all extracted spectra are shown in Table \ref{table:individual}, where the uncertainties correspond to the 90\% confidence level. 
\par
All regions require two thermal components, indicating that the plasma of Kes 79 contains a soft and hard temperature component. The temperatures derived for the hard component in regions 3 - 14 range between $0.61 \pm 0.03$ - $1.36 \pm 0.3$ keV, while for the filamentary structures encompassed by regions 1 - 2, the hard temperatures are equal to  $0.55^{+0.03}_{-0.01}$ keV and $0.49 \pm 0.03$ keV respectively.  The derived VPSHOCK temperatures for regions 3 - 14 region are, on average, comparable to the temperature derived by \citet{2004ApJ...605..742S} and \citet{2009A&A...507..841G}, who used a single component NEI model, while the temperature of the filamentary structures are lower than global temperatures derived by \citet{2004ApJ...605..742S} and \citet{2009A&A...507..841G}. The soft plasma (PSHOCK) component of Kes 79's X-ray emitting material is highly isothermal in nature, with all regions requiring a temperature of $\sim 0.24$ keV. 
\par
All region-specific spectra are similar to each other, with strong emission lines coming from Si, S, and Mg, with some requiring Ne and Ar.  All regions in the high kT component have super-solar abundances (e.g., larger than $\sim$ 2 times solar abundance) of Si and S, while Mg has $\sim$ 1.5 times solar abundance. We thus associate this component with shocked ejecta. All regions require approximately solar abundances of Ne, with the exception of region 7 which has an enhancement of Ne. Region 15 has the largest elemental abundances out of all the regions analysed, with Mg, Si and S requiring super solar abundances. The inner regions closest to the X-ray position of the pulsar (region 7, 11 and 12) and the filament described by region 2 all require the highest abundances of Mg, Si, and S overall.  All regions require enhanced abundances of Mg, Si and S, which indicates that the ejecta component is found everywhere through out the remnant, implying that we observed X-ray emission from shocked ejecta. The abundances obtained in this analysis are larger than the values obtained by \citet{2004ApJ...605..742S} using their ionisation equilibrium collisional plasma model or non-equilibrium ionisation collisional plasma models. This arises from the fact that in our analysis, we use the abundance table by \citet{2000ApJ...542..914W}, while \citet{2004ApJ...605..742S} uses the table derived by \citet{1989GeCoA..53..197A}. 
\par
The shocked ejecta, modeled using VPSHOCK, have an ionisation timescale of $\sim 10^{11} $ cm$^{-3}$s in all regions. These values agree well with the ionisation timescales derived by \citet{2004ApJ...605..742S} and \citet{2009A&A...507..841G}. The ionisation timescale of the material swept up by the forward shock as derived using the PSHOCK model is fixed to $1 \times 10^{12} $ cm$^{-3}$s, as initial spectral analysis yielded ionisation timescales of this magnitude. The high ionisation timescale of all regions indicates that the soft plasma component is in ionisation equilibrium \citep{2012A&ARv..20...49V}, while the hard plasma component across the whole remnant is far from ionisation equilibrium. 
\par
The derived hydrogen column density for each region ranges between $(2.08^{+0.03}_{-0.04} - 2.75^{+0.03}_{-0.04}) \times 10^{22}$ cm$^{2}$. These values are higher than the estimated absorbing column density derived by \citet{2004ApJ...605..742S} using \textit{Chandra} data ($\sim 1.6 \times 10^{22}$ cm$^{2}$) and \citet{2009A&A...507..841G} using \textit{XMM-Newton} data ($\sim 1.5 \times 10^{22}$ cm$^{2}$). This discrepancy most likely arises from the fact that we use a different abundance table in our analysis and a two component plasma model.  These derived values indicate that the X-ray spectrum of Kes 79 is heavily absorbed by interstellar material.  

\subsubsection{Global spectral properties}
\begin{figure}[t!]
\includegraphics[width=\textwidth]{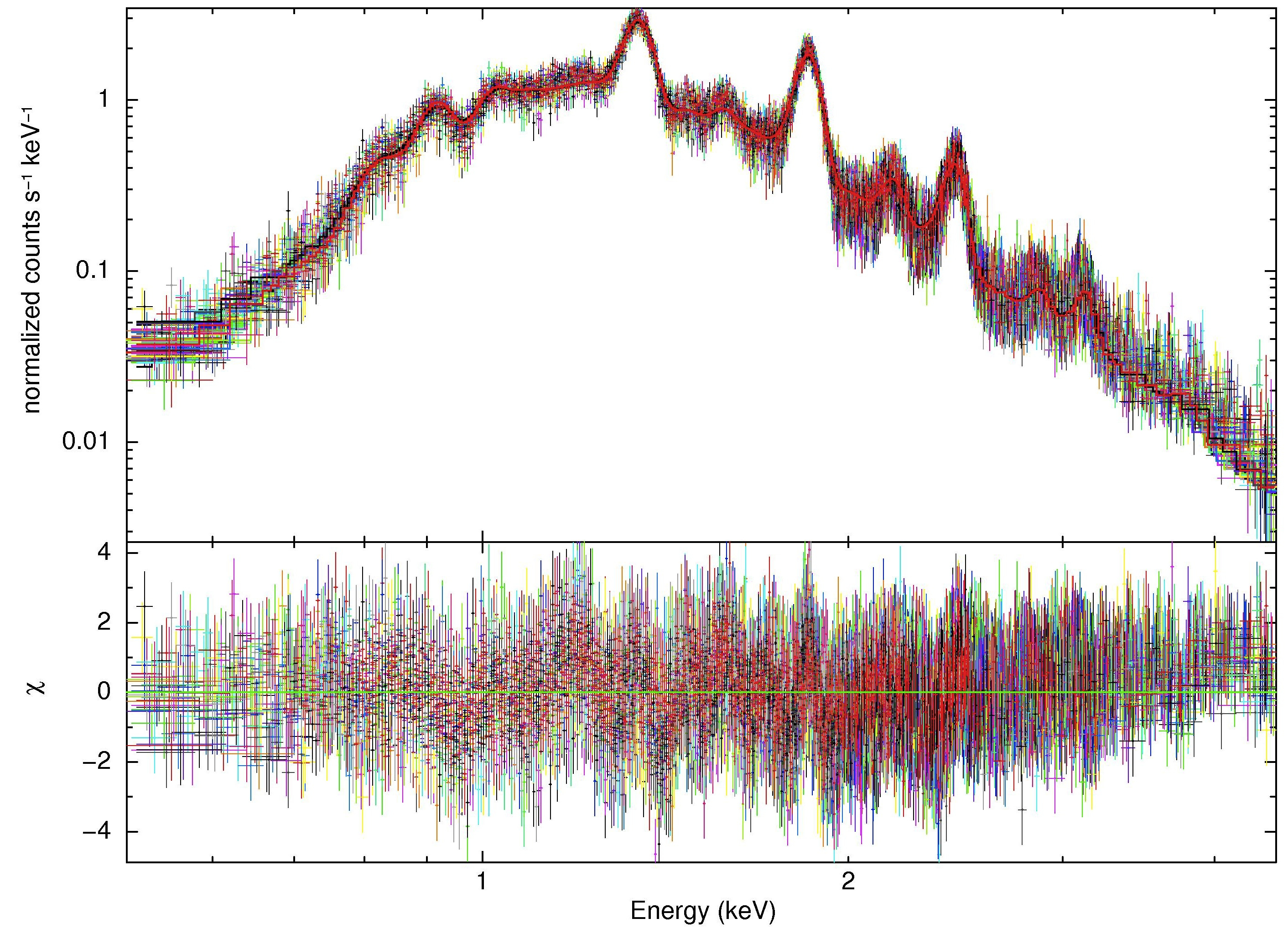}
\caption{The MOS 1 and MOS 2 spectrum extracted from 16 $XMM-Newton$ observations with good time intervals of 10 ks or more for the whole of Kes 79. These spectra have been fitted simultaneously with a two component plasma model, VPSHOCK+PSHOCK, and the XSPEC absorption model TBABS described by the parameters in Table \ref{table:global}. The X-ray spectra are overlaid with the fitted VPSHOCK+PSHOCK model and their chi-square residuals are shown. The combined MOS 1 and MOS 2 X-ray spectra derived for each region shown in Figure \ref{fig:xraystuff}a all have similar features as the global X-ray spectrum shown here.  \label{figure:global}}
\end{figure}

To investigate the average spectral properties of Kes 79, we extracted a global X-ray spectrum from a circular region centered at  ($\alpha_{J2000}$, $\delta_{J2000}$)=($18^{h}52^{m}41.31^{s}, +0^{\circ}40^{'}52.97^{"}$) with a radius of 4.70 arcminutes. This region enclosed the complete SNR as observed in X-rays. The background spectrum was extracted using the four source-free regions shown in Figure \ref{fig:xraystuff}a. We did this for each observation with good exposure time of 10 ks or greater and all spectra are binned with a minimum of 25 counts. We fit all 16 MOS1 and MOS2 observations simultaneously over an energy range of 0.5 - 10 keV. Similar to the spatial X-ray analysis, a single component VPSHOCK model with enhanced abundances of Si, S and Mg produced unsatisfactory fits, and was improved when we added a thermal component (PSHOCK) ($\chi^{2}/(dof) = 1.07$). Figure \ref{figure:global} shows the global X-ray spectrum of Kes 79, while the values of these best fit parameters with their corresponding one sigma uncertainties are found in Table \ref{table:global}. This result implies that the plasma of Kes 79 consists of a hard component with a plasma temperature of $0.77^{+0.006}_{-0.001}$ keV and a soft component with a temperature of $0.24^{+0.0002}_{-0.001}$ keV. This is consistent with our fits of the individual regions. The temperatures derived by \citet{2004ApJ...605..742S}  and \citet{2009A&A...507..841G} using \textit{Chandra} and \textit{XMM} data is similar to the temperature of the hard plasma component. The best-fit absorbing column density is $N_{H} =2.35^{+0.003}_{-0.01} \times 10^{22}$ cm$^{2}$ and is higher than values obtained by \citet{2004ApJ...605..742S} and \citet{2009A&A...507..841G} using a single NEI plasma model. Using the Anders and Grevesse abundance table we obtained a column density similar to that obtained by \citet{2004ApJ...605..742S} and \citet{2009A&A...507..841G}.
\par
Using the best fit of the global X-ray spectrum and the total count rate, we derive an unabsorbed flux, for an energy range of 0.5 - 10 keV, of $9.8 \times 10^{-10}$ erg cm$^{-2}$ s$^{-1}$ and a luminosity of $5.6 \times 10^{36}$ erg s$^{-1}$ assuming a distance of 7 kpc. The luminosity derived in this analysis is a factor of 2 larger than the flux obtained by \citet{2004ApJ...605..742S} using \textit{Chandra} and this most likely arises from the fact we use a two component NEI plasma model, while \citet{2004ApJ...605..742S} uses a one component NEI model for their analysis.
%
%

\begin{table}[t!]
\caption{Results from fitting the two component model VPSHOCK+PSHOCK model to the whole SNR with the 90\% confidence ranges. \label{table:global}}
\begin{small}
\begin{center}
\begin{tabular}{cc}
\hline
Parameters& Global fit\\
\hline
nH($10^{22}$  cm$^{2}$)&$2.35^{+0.003}_{-0.01} $ \\
\textit{Soft Component (PSHOCK)}&\\
kT (keV)&$0.24^{+0.0002}_{-0.001}$ \\
$\tau (10^{12}$ s cm$^{-3})$&$1$ \\
Norm (cm$^{-5}$)&$0.67^{+0.0002}_{-0.001}$ \\
Absorbed flux for 0.5 - 10 keV (erg cm$^{-2}$ s$^{-1}$) & $2.65 \times 10^{-12}$ \\
Unabsorbed flux for 0.5 - 10 keV (erg cm$^{-2}$ s$^{-1}$) & $7.68 \times 10^{-10}$ \\
\textit{Hard Component (VPSHOCK)}&\\
kT (eV)&$0.78^{+0.006}_{-0.001}$ \\
$\tau (10^{11}$ s cm$^{-3})$ & $1.57 \pm 0.03$ \\
Norm $cm^{-5}$&$0.06^{+0.01}_{-0.02}$ \\
Mg&$1.41^{+0.02}_{-0.01}$ \\
Si&$2.33 \pm 0.02$ \\
S&$2.28 \pm 0.03$ \\
Absorbed flux for 0.5 - 10 keV (erg cm$^{-2}$ s$^{-1}$) & $7.90 \times 10^{-12}$ erg cm$^{-2}$ s$^{-1}$\\
Unabsorbed flux for 0.5 - 10 keV (erg cm$^{-2}$ s$^{-1}$)  & $2.09 \times 10^{-10}$ erg cm$^{-2}$ s$^{-1}$\\
$\chi^{2}/(dof)$&1.07\\
Total luminosity (0.5 - 10 keV) & $ 5.6\times 10^{36}$ erg s$^{-1}$\\
\hline
\end{tabular}
\end{center}
\end{small}
\vspace{0.5cm}
\end{table}


\subsubsection{X-ray characteristics of Kes 79}
Using the PSHOCK+VPSHOCK model parameters summarised in Table \ref{table:individual}, we derive the density of the X-ray emitting material and how it changes across the remnant. In our calculations we set the radius (R) of Kes 79 to $4.7$ arcmin, which corresponds to a physical size of 9.6 $d_{7}$ pc. For each individual region, the X-ray emitting volume was estimated by taking an area equivalent to the extracted SNR regions (Figure \ref{fig:xraystuff}a) and projecting this area through a shell (front and back) with a thickness of R/12.  The volume emission measure ($EM$), which describes the amount of plasma available to produce the observed X-ray flux, is defined as $EM \sim \int n_{e} n_{H} dV_{X} \sim n_{e} n_{H} f V_{X} $, where $n_{e}$ is the post-shock electron density, $n_{H}$ is the mean hydrogen density and $f$ is a volume filling factor . The $EM$ is estimated using the normalisation of the X-ray spectral fits using $K=(10^{-14}/4\pi d^{2}) EM$. Assuming a strong shock, the ambient density can be calculated using $n^{2}_{0}=(K d^{2} \pi)/(4.8\times10^{-14} f V_{X})$. From this, we can calculate the ambient density ($n_{0}$) across the remnant (i.e. for each region). These $n_{0}$ calculations are summarised in Table \ref{table:regionambient}.  The inferred ambient density across the remnant ranges from $n_{0} = (1.81 \pm 0.5 - 4.85^{+1.1}_{-0.9}) f^{-1/2}d_{7}^{-1/2} cm^{-3}$. These high ambient density values indicate that the remnant is expanding into a dense environment. The ambient density is highest in regions 5, 6, 9, 10, 11 and 16 which are predominantly towards the south and southeast of the remnant. This observation is consistent with the presence of a nearby molecular cloud overlapping the eastern and southeastern regions of Kes 79. The presence of X-ray emission in these regions may imply the interaction of the remnant's shock-wave with the molecular cloud.
\par
Our estimates of the ambient density are an order of magnitude larger than the estimated ambient density by \citet{2004ApJ...605..742S}. This results from the higher column density derived using the \citet{2000ApJ...542..914W} abundance table. This absorption hides a significant soft component, which comprises a large total mass and results in a high derived density. The temperature derived by \citet{2004ApJ...605..742S} using the three non-ionisation equilibrium models in their paper have approximately intermediate values of the soft and hard temperature components derived from our spatial dependent and global X-ray spectra (Table \ref{table:individual} and Table \ref{table:global}, respectively) . This could imply that the  short \textit{Chandra} observation could not distinguish between the two plasma components suggested by our extracted X-ray spectra and instead represents an average of the soft and hard components summed along the line of sight.
\par
In the Sedov phase of evolution, the SNR radius scales as $R \propto n_{0}^{-1/5}$, at a given age. Thus, while our analysis indicates a somewhat non-uniform ambient medium, the overall behaviour can be reasonably approximated by assuming evolution in a uniform medium \citep{1999A&A...344..295H}. When calculating the global characteristics of Kes 79, we estimate the volume of the X-ray emitting region ($V_{X}$) by assuming that the plasma fills a spherical shell with a radius of 9.6 $d_{7}$ pc,  a thickness of 2 times R/12 and the X-ray parameters listed in Table \ref{table:global}. The shock temperature can be derived by using: $T_{sh} = 0.79 T_{X}$, where $T_{X}$ is the temperature inferred from the X-ray fit, assuming electron-ion temperature equilibrium. Using the global temperature for the soft component (PSHOCK) which is suggested to be the forward shock component, we obtain a shock temperature of $T_{sh} \sim 0.19$ keV. The forward shock velocity is determined using $v_{sh} = [(16 k T_{sh})/(3 \mu m_{H})]^{1/2}$, where $\mu = 0.604$ is the mean atomic weight, $k$ is Boltzmann's constant and $m_{H}$ is the mass of hydrogen. We obtain a shock velocity of $v_{sh} \sim  400$ km s$^{-1}$, assuming electron-ion temperature equilibration. The Sedov age of the remnant is related to the forward shock velocity by $t_{SNR} = 2r_{sh}/5v_{sh}$. Assuming the SNR radius of $r_{sh}=4.7$ arcminutes, we derive an age of $\sim 9400 d_{7}$ years. Our spatial X-ray analysis described by the model parameters in Table \ref{table:individual} gives a $t_{age}$ between $\sim 9000 - 10600$ years whose average is approximately the $t_{age}$ that we derived in the global analysis. \citet{2004ApJ...605..742S}, using different X-ray model parameters, derived an age of Kes 79 of 5400 - 7500 years, while \citet{2009A&A...507..841G} using VLA radio data derived an upper limit of $t_{age} \textless 15 \times 10^{3}$ years. The explosion energy is calculated using $E_{SNR} = 0.7 r_{sh}^{5}m_{p}n_{0}t_{SNR}^{-2}$ and we obtain $E_{SNR} \sim 3.7\times10^{50}f^{-1/2}d_{7}^{-5/2}$ erg. This value is lower than the canonical $10^{51}$ erg. This low value for the explosion energy has been inferred from X-ray observations of a number of other SNRs, e.g., G272.2-3.2 \citep{2001ApJ...552..614H} and G299.2-2.9 \citep{2007ApJ...665.1173P}. \citet{2011ApJ...734...85C}, suggested that such low inferred explosion energies could be a signature of efficient cosmic ray acceleration during the evolution of the SNR, consistent with the detection of $\gamma$-rays from Kes 79 (see below). The X-ray emitting mass is calculated using $M_{X} = 1.4 n_{H} m_{H} f V$ and we obtain $M_{X} \sim 160 f^{1/2} d^{5/2}_{7.1} M_{\odot}$ of swept-up ISM.


\begin{table}[t!]
\caption{Derived ambient density estimates for the 16 different regions shown in Figure \ref{fig:xraystuff}a. These values are calculated from the best fit parameters of our two component plasma model, VPSHOCK+PSHOCK, and the XSPEC absorption model TBABS shown in Table \ref{table:individual}. \label{table:regionambient}}
\begin{small}
\begin{center}
\begin{tabular}{cc}
\hline
Region&  Ambient Density  \\
&  ($d_{7} f^{-0.5} cm^{-3}$)\\
\hline
 1 & $3.39^{+2.2}_{-2.0}$ \\
 2 & $2.52^{+2.2}_{-1.6}$ \\
 3 & $2.49^{+1.4}_{-0.6}$ \\
 4 & $3.15^{+0.9}_{-1.0}$ \\
 5 & $4.29^{+2.3}_{-1.7}$ \\
 6 & $3.59^{+1.6}_{-1.2}$ \\
 7 & $3.08^{+1.0}_{-1.2}$ \\
 8 & $2.68^{+1.1}_{-1.4}$ \\
 9 & $3.95 \pm 0.6$ \\
 10 & $4.85^{+1.1}_{-0.9}$ \\
 11 & $4.29^{+1.3}_{-2.3}$ \\
 12 & $3.94^{+2.3}_{-1.8}$ \\
 13 & $2.34^{+1.7}_{-1.1}$ \\
 14 & $2.55 \pm 1.5$ \\
 15 & $1.81 \pm 0.5$ \\
 16 & $3.51^{+2.5}_{-2.0}$ \\
\hline
\end{tabular}
\end{center}
\end{small}
\vspace{0.5cm}
\end{table}


\section{DISCUSSION}

Non-thermal X-ray and $\gamma$-ray emission from SNRs indicate that a population of particles are being accelerated to very high energies by the SNR shock front. $\gamma$-rays can be produced by leptonic mechanisms such as inverse Compton scattering, non-thermal bremsstrahlung, or by hadronic mechanisms via the decay of neutral pions generated during proton-proton interactions. SNR shocks are not the only astrophysical source capable of producing $\gamma$-rays. Pulsars can also contribute to the observed $\gamma$-ray emission and their contribution should also be considered before we assume that the $\gamma$-ray sources are products of particles being accelerated at the SNR shock.

\subsection{Pulsar contribution to $\gamma$-rays }
In the first $Fermi$-LAT Pulsar catalogue, \citet{2010ApJS..187..460A} determined that the spectra of pulsars detected in the $Fermi$-LAT energy band are best described by a power-law distribution with an exponential cut-off of 1 - 5 GeV, with the exception of the Crab pulsar which can be described using an exponential cut-off of $\textgreater$100 GeV \citep{2011Sci...334...69V}. This result was confirmed and presented in the second $Fermi$-LAT Pulsar catalogue \citep{2013arXiv1305.4385T} in which the $Fermi$-LAT collaboration analysed three years of $Fermi$-LAT data and detected with high confidence, 117 $\gamma$-ray pulsars above an energy of 0.1 GeV . Most recently, a preliminary analysis by \citet{2012AIPC.1505..293S} of pulsars detected by the $Fermi$-LAT satellite has indicated that a number of pulsars show possible evidence for high energy emission with a cut-off energy $\textgreater$ 10 GeV. The spectra of the source coincident with Kes 79 can be best fit using an exponential cut-off of $E_{cut} \sim$ 2.62 GeV  and $E_{cut} \sim$ 1.5 GeV respectively, which falls within the expected range for $E_{cut}$ for $\gamma$-ray pulsars. As a consequence, the hypothesis that the observed $\gamma$-ray emission could be due to the presence of a pulsar cannot be ruled out straight away.
\par
Within a 5$^{\circ}$ region centered on the observed $\gamma$-ray emission of Kes 79 (Figure \ref{fig:count}a), there are three pulsars in the Australian Telescope National Facility (ATNF) Pulsar Catalogue \citep{2005AJ....129.1993M}\footnote{The ATNF Catalogue contains the positional and physical characteristics of 1509 pulsars which can be found at \url{http://www.atnf.csiro.au/research/pulsar/psrcat}}, whose spin down power is sufficient to produce the $\gamma$-ray flux of Kes 79 if, at the distance of the pulsar, all of that power goes into $\gamma$-rays. An upper limit on the $\gamma$-ray flux of each ATNF pulsar is calculated by using the distance to the pulsar listed in the catalogue and by assuming that 100 \% of the spin down power, $\dot{E}$, is converted into $\gamma$-rays. As the upper limit on the pulsar's $\gamma$-ray flux is comparable to the observed $\gamma$-ray flux of Kes 79, this indicates that these three pulsars could potentially contribute to the observed $\gamma$-rays. The pulsars would also have to be located such that the \textit{Fermi}-LAT's PSF could not resolve them as separate sources. At 0.6 GeV, which corresponds to the energy of the first data point of Kes 79's $\gamma$-ray spectrum, the resolution of the PSF of the \textit{Fermi}-LAT is $\sim 1^{\circ}$ for front events at 68\% containment\footnote{\url{$http://www.slac.stanford.edu/exp/glast/groups/canda/lat\_performance.htm$}}. All three pulsars are at a distance of 1.23$^{\circ}$ or more from the centroid of the $\gamma$-ray emission, thus making it highly unlikely that any of these pulsars are contributing significantly to the observed $\gamma$-ray emission of Kes 79. The 105 ms X-ray pulsar, PSR J1852+00040 discovered by \citet{2005ApJ...627..390G} and \citet{2003ApJ...584..414S} to be associated with Kes 79, has a spin down power $\dot{E}$ =  3$\times$10$^{32}$ erg s$^{1}$ \citep{2010ApJ...709..436H}. If all of this spin down energy is converted into $\gamma$-rays at the distance of the pulsar, it is incapable of producing the $\gamma$-ray flux we observe. Thus, we can state with confidence that this pulsar is not producing the observed $\gamma$-ray emission of Kes 79.

\subsection{Hadronic origin of $\gamma$-rays}
As there is a lack of evidence in favour of significant $\gamma$-ray contribution from a pulsar, it is reasonable to assume that the observed $\gamma$-rays associated with Kes 79 arise from the SNR itself, a prospect made particularly likely by the high ambient density implied by our X-ray analysis in Section 3. This is supported by the presence of strong HCO$^{+}$ J=1 $\rightarrow$ 0 and $^{12}$CO J=1 $\rightarrow$ 0 emission found towards the east of the remnant and the fact that the remnant is thought to be interacting with dense molecular material \citep{1987ApJS...63..821S, 1989MNRAS.238..737G}. In order to investigate the nature of the $\gamma$-ray emission, we first model the $\gamma$-ray spectrum assuming that the $\gamma$-rays are produced predominantly by the decay of accelerated hadrons. 
\par
To reproduce the observed $\gamma$-ray spectrum of Kes 79, we use a model which calculates simultaneously the $\gamma$-ray emission assuming a distribution of electrons and protons being accelerated at the SNR shock front.  The $\pi^{0}$ decay model is based on a proton-proton interaction model by \citet{2006ApJ...647..692K}, with a scaling factor of 1.85 for helium and heavy nuclei \citep{2009APh....31..341M}. The synchrotron and inverse Compton emission models are based on \citet{1999ApJ...513..311B} and the non-thermal bremsstrahlung emission is modeled using the description in \citet{2000ApJ...538..203B}. To generate the momentum distribution of protons (or electrons) being accelerated by the SNR shock wave, we adopt a power law spectrum with an exponential cut-off, as expected from classical diffusive shock theory. This is described by: 
\begin{equation}\label{eq:uno}
\frac{dN_{i}}{dp} = a_{i} \,p^{-\alpha_{i}} \exp\left(-\frac{p}{p_{0\,i}}\right),
\end{equation}
where $i$ is the particle species, $a_{i}$ is the normalisation of the particle distribution and $p_{0\,i}$ is the exponential particle momentum cut-off. This particle distribution is transformed from momentum space to energy space, such that the exponential cut-off is defined by an input energy, $E_{0i}$. The integrals of each particle species distribution are summed and set such that the sum is equal to the assumed total energy in accelerated particles, $E_{CR}$ = $\epsilon E_{SN}$, where $\epsilon$ is the efficiency of the SNR in depositing energy into cosmic rays. 
\par
\begin{figure}
\includegraphics[width=\textwidth]{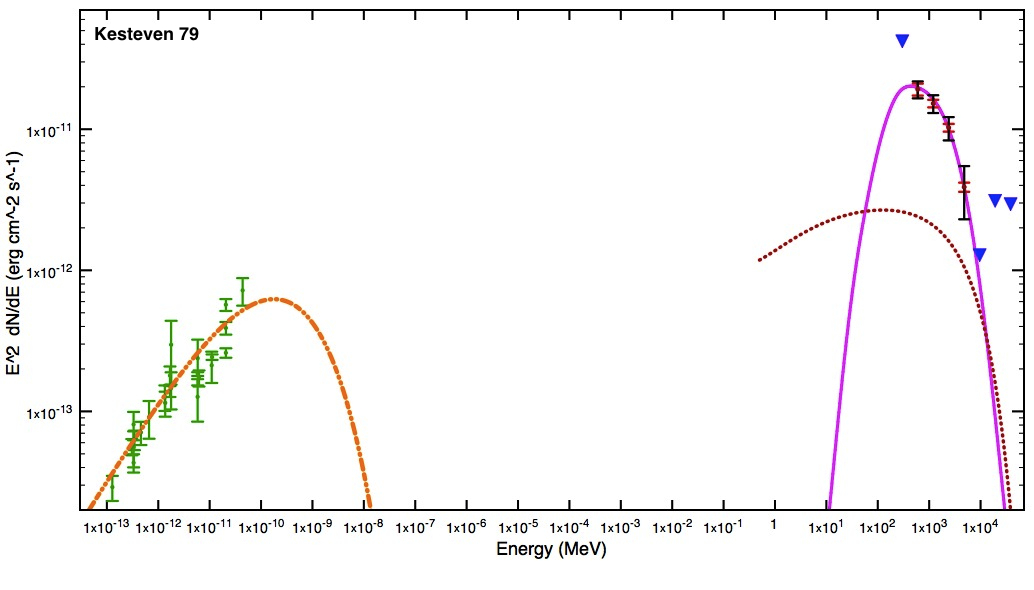}
\caption{The broadband fits to radio (green data points) and $Fermi$-LAT $\gamma$-ray data points (as described in Figure \ref{fig:gamma}) of Kes 79. The pion decay, non-thermal bremstrahlung and synchrotron models defined by the parameters in Table \ref{Table:model} are shown as the solid magenta line, dotted dark red line and dot-dashed orange line, respectively. The corresponding IC model falls below the plotted axis. \label{fig:broadband}}
\end{figure}

Figure \ref{fig:broadband} shows the model fits to the broadband emission observed from Kes 79.  The radio spectrum of Kes 79, shown as the green data points in Figure \ref{fig:broadband}a, is a combination of multiple observations by \citet{1967AnAp...30..723K, 1968AuJPh..21..369K, 1969AuJPh..22..121B, 1970A&AS....1..319A, 1972AuJPh..25..429D, 1973AuJPh..26..369D, 1975A&A....45..239C, 1975AJ.....80..679B, 1975AJ.....80..437D, 1977AuJPA..43....1S, 1977A&A....55...11A, 1981MNRAS.195...89C, 1989ApJS...71..799K, 1992AJ....103..943K, 1994ARep...38...95K, 2009A&A...507..841G}.
We have plotted the pion decay model that adequately reproduces the observed $\gamma$-ray spectrum of Kes 79 as the solid magenta line, while the orange dot-dashed line corresponds to the synchrotron model which sufficiently reproduces the radio spectrum of each remnant, assuming an electron-positron ratio ($K_{ep}$) of 0.01. For completeness we have also plotted as the dotted dark red line, the non-thermal bremstrahlung contribution for Kes 79, while the corresponding IC model falls below the plotted axis. In Table \ref{Table:model} we have listed the parameters which reproduce the pion-decay, synchrotron, non-thermal bremsstrahlung and IC models observed in Figure \ref{fig:broadband}.
\par
The $\gamma$-ray spectrum of Kes 79 is adequately described by a pion-decay model with a proton distribution that has a power law index of $\alpha_{p} = 4.09$ and a proton cut-off energy of $E^{p}_{0} \sim 12 $ GeV. 
The cut-off energy for the proton spectrum required to model the $\gamma$-ray spectrum of Kes 79 under this hadronic-dominated scenario is rather low given that we expect cut-off energies for protons to be well in excess of a TeV \citep{2008ARA&A..46...89R}. This discrepancy could arise from efficient CR acceleration being suppressed in dense surrounding environments due to the suppression of Alfven waves by ion-neutral collisions. This allows particles being accelerated by the SNR shock to escape the emission volume \citep{2011NatCo...2E.194M, 2012PhPl...19h2901M}.
\par
To model the radio spectrum of Kes 79 using synchrotron emission, we assume that the cut-off energy of the electrons to be the same as the cut-off energy of the pion-decay model, as there is no non-thermal X-ray data to constrain the cut-off energy of the electron population. To reproduce the radio spectrum of Kes 79, we required an electron distribution with a power law index of $\alpha_{e} = 3.8$ and a magnetic field of 180 $\mu$G.
The magnetic field required to model the radio spectrum of Kes 79 is larger than the average interstellar magnetic field (3 - 5 $\mu$G) and could arise from the magnetic fields being amplified by the shock-wave compressing the surrounding medium during the acceleration process. 
\par
If one makes reasonable assumptions about the fraction, $\epsilon$, of total supernova explosion energy, $E_{SN}$, used to accelerated cosmic rays, as well as the shock compression ratio, $r$, one can use this model to estimate the density of the surrounding medium that the SNR shock wave is interacting with. For this analysis we set $r=4$ and we assume that $\epsilon$ = 0.4. This is a conservatively high estimate of the amount of supernova explosion energy expected to be converted into accelerating cosmic rays. In Table \ref{Table:model} we summarise the estimated density of the $\gamma$-ray emitting material as well as the density of the X-ray emitting material derived in the literature and from our X-ray analysis.
\par
When fitting the $\gamma$-ray spectrum of Kes 79 using a pion-decay model, we obtained a density of 14.5 cm$^{-3}$. This density is much larger than the values derived from X-ray measurements, listed in Table \ref{table:regionambient}. This inferred density enhancement could indicate that the shock-wave of the SNR is interacting with cold, dense clumps of material (ISM or ejecta) that form due to instabilities in the post-shock flow \citep{2010ApJ...717..372C, 2012ApJ...744...71I}. These cold dense clumps do not radiate significantly in X-rays. If these clumps of material have a high filling factor, then the densities inferred from the X-ray emission would significantly underestimate the local mean density. This scenario is consistent with a CO survey conducted by \citet{1987ApJS...63..821S} which revealed a large elongated molecular cloud that overlaps the eastern and south eastern region of Kes 79.  Molecular clouds are known to be clumpy in nature, with dense clumps of molecular material occupying only a fraction of the actual molecular cloud volume. This clumpy material can have a density of $n_{H} \sim 5 - 25$ atoms cm$^{-3}$ \citep{1999ApJ...511..798C} or an ambient density of $n_{0} \sim 1.25 - 6.25$ cm$^{-3}$, if we assume $n_{0}=n_{H}/4$. Our inferred ambient densities (Table \ref{table:regionambient}) across the remnant fall within this range, supporting the original conclusion that Kes 79 is interacting with a dense molecular material that does not radiate significantly in X-rays.
\par
Alternatively, highly energetic particles accelerated by the remnant's shock-wave could escape the acceleration region and stream ahead of the shock. These escaped particles can interact with the dense gas upstream of the shock, enhancing the detected $\gamma$-ray emission. A number of authors such as \citet{1996A&A...309..917A}, \citet{2008AIPC.1085..265G} and \citet{2009ApJ...707L.179F} have shown that dense molecular clouds in the vicinity of a supernova remnant can significantly influence the $\gamma$-ray flux. However, the escaping particles in this scenario come predominantly from the higher energy part of the particle spectrum, leading to inconsistencies between the predicted and observed $\gamma$-ray spectrum.

\begin{table}[t!]
\begin{center}
\caption{Model parameters and density estimates for the pion decay model for Kesteven 79. \label{Table:model} }
\begin{tabular}{cccccccccccc}
\hline
Object & Distance &$\alpha_{proton}$ &$\alpha_{elec}$ &$E_{0}^{proton}$&$E_{0}^{elec}$ &Magnetic field& Ambient density & X-ray density & $n_{X}$ reference \\
            &(kpc)       &&  & (GeV) & (GeV) &($\mu G$)	& $n_{0}$ (cm$^{-3}$) 	& $n_{X}$ (cm$^{-3}$) 	& \\
\hline
Kes 79 	  &7& 4.09 & 3.8& 12&12 &180 &14.5 & $4^{+0.9}_{-0.7}$ $d_{7} f^{-0.5}$ & Table \ref{table:regionambient}\\
\hline
\end{tabular}
\end{center}
\vspace{0.5cm}
\end{table}
\subsection{Leptonic origin of $\gamma$-rays}

An alternative scenario is that the $\gamma$-ray emission coincident with Kes 79 arises from a non-thermal population of electrons in the form of IC or bremsstrahlung radiation.  \citet{2000ApJ...538..203B} and other authors, have proposed this mechanism as the main production mechanism of $\gamma$-rays in SNRs. For completeness, we tested whether IC or non-thermal bremsstrahlung could be the dominant mechanism producing the $\gamma$-ray spectra of Kes 79.  If one wants non-thermal bremsstrahlung to dominate the GeV $\gamma$-ray emission we need to have the electron-to-proton ratio, $K_{ep}$, to be larger than $\sim 0.2$ \citep{1998ApJ...492..219G}. Local observations of the cosmic-ray abundance ratio imply that $K_{ep} \sim 0.01$ \citep{1998ApJ...492..219G}, while models of the $\gamma$-ray emission from SNRs such as RX J1713.7-3946 suggest even smaller values of this ratio \citep{2010ApJ...712..287E}.  
\par
Assuming that electrons make up approximately 1\% of the relativistic particle energy budget (i.e.  $K_{ep}$ = 0.01), we determined that in order to reproduce the $\gamma$-ray spectrum of Kes 79, the total electron energy would be $1 \times 10^{51}$ erg.  This is equal to the entire canonical SN explosion energy, making it difficult to assert that IC is the dominant mechanism behind the $\gamma$-ray emission of Kes 79. 
\par
To test whether non-thermal bremsstrahlung is the dominant mechanism producing the $\gamma$-rays of Kes 79, we used the highest ambient density value derived in our X-ray analysis (see Table \ref{table:regionambient}), as this would give us a lower limit on the total energy in accelerated electrons. Assuming $K_{ep}$ = 0.01, we obtained a lower limit of $6.1 \times 10^{50}$ erg for the total electron energy. This implies a total relativistic particle energy far in excess of the mechanical energy of a SNR.  This fact makes it unlikely that non-thermal bremsstrahlung is the dominant mechanism behind the observed $\gamma$-rays of Kes 79. 

\section{Conclusion}

In summary, using 52 months of \textit{Fermi}-LAT data we searched for $\gamma$-ray emission from SNR Kes 79, which is found to be interacting with molecular clouds. Assuming that the main production mechanism for the observed $\gamma$-ray emission arises from pion decay, the inferred ambient density is high. This is consistent with observations of dense molecular clouds near Kes 79. However, these estimates are larger than the density derived from X-ray observations found in the literature. A pulsar association with the observed $\gamma$-ray emission is also considered for completeness as well as leptonic origin of the $\gamma$-ray emission. For Kes 79, SNR origin of the emission is the most likely scenario, while hadronic origin of the $\gamma$-rays was the most energetically favoured mechanism. Additionally we performed a spatial and spectral analysis of Kes 79 using 16 archival \textit{XMM-Newton} observations and inferred an ambient density that is much smaller than the inferred ambient density from modeling the observed $\gamma$-ray emission. These results are consistent with scenarios represented by \citet{2010ApJ...717..372C} and \citet{2009ApJ...706L...1A}. We also performed a similar analysis for SNR Kes 78, but due to the uncertainities in the background $\gamma$-ray model or from $\gamma$-ray emission spilling over from nearby sources, we were unable to concluded that there is an excess of GeV $\gamma$-ray emission associated with this remnant.
\par
The authors would like to thank Dr. Thomas M. Dame for his in-depth discussions in regards to the association and interaction of Kes 79 with its surrounding molecular cloud. KA would like to thank Dr. Jasmina Lazendic-Galloway for her helpful discussions in regards to the X-ray analysis. 
This work was partially funded by NASA grand NNX 11AQ096. PS acknowledges support from NASA contract NAS8-03060. DC acknowledges support for this work provided by NASA through SAO contract SV3-73016 to MIT for Support of the Chandra X-Ray Center, which is operated by the SAO under contract NAS8-03060.


\end{document}